\documentclass[journal,twoside,web]{ieeecolor}
\usepackage{tmi}
\usepackage{cite}
\usepackage{amsmath,amssymb,amsfonts}
\usepackage{algorithmic}
\usepackage{graphicx}
\usepackage{textcomp}
\usepackage{booktabs}
\usepackage{multirow}

\def\BibTeX{{\rm B\kern-.05em{\sc i\kern-.025em b}\kern-.08em
    T\kern-.1667em\lower.7ex\hbox{E}\kern-.125emX}}
    
\begin{document}
\title{AFFIRM: Affinity Fusion-based Framework for Iteratively Random Motion correction of multi-slice fetal brain MRI}
\author{Wen Shi, Haoan Xu, Cong Sun, Jiwei Sun, Yamin Li, Xinyi Xu, Tianshu Zheng, Yi Zhang, Guangbin Wang and Dan Wu

\thanks{This study was supported by the Ministry of Science and Technology of the People’s Republic of China (2018YFE0114600), National Natural Science Foundation of China (61801424, 81971606, and 82122032), the Leading Innovation and Entrepreneurship Team of Zhejiang Province (2020R01003 and 2022C03057), and the Young Scientist Program of United Imaging (UIH-QNJJ-2021001). The authors thank Dr. Michael Ebner from King's College London for constructive help on NiftyMIC toolbox. (Corresponding author: Dan Wu)}
\thanks{W. Shi, H. Xu, J. Sun, X. Xu, T. Zheng, Y. Zhang, D. Wu are with Key Laboratory for Biomedical Engineering of Ministry of Education, Department of Biomedical Engineering, College of Biomedical Engineering \& Instrument Science, Zhejiang University, Hangzhou, 310027, China. (email: wshi15@jhmi.edu; 22115011@zju.edu.cn; 3170105381@zju.edu.cn; xuxinyi\_bme@zju.edu.cn; zhengtianshu@zju.edu.cn; yizhangzju@zju.edu.cn; danwu.bme@zju.edu.cn).}
\thanks{C. Sun, G. Wang are with Department of Radiology, Shandong Medical Imaging Research Institute, Cheeloo College of Medicine, Shandong University, Jinan, 250021, China (email: suncong\_jida1991@163.com; wgb7932596@hotmail.com).}
\thanks{Y. Li is with School of Biomedical Engineering, Shanghai Jiao Tong University, Shanghai, 200240, China (email: yaminli19@sjtu.edu.cn).}
\thanks{W. Shi is also with Department of Biomedical Engineering, Johns Hopkins University School of Medicine, Baltimore, MD, 21287, USA.}
\thanks{D. Wu is also with Center for Intelligent Biomedical Instrumentation, Zhejiang University Binjiang Research Institute, Hangzhou, 310052, China.}}

\maketitle

\begin{abstract}
Multi-slice magnetic resonance images of the fetal brain are usually contaminated by severe and arbitrary fetal and maternal motion. Hence, stable and robust motion correction is necessary to reconstruct high-resolution 3D fetal brain volume for clinical diagnosis and quantitative analysis. However, the conventional registration-based correction has a limited capture range and is insufficient for detecting relatively large motions. Here, we present a novel Affinity Fusion-based Framework for Iteratively Random Motion (AFFIRM) correction of the multi-slice fetal brain MRI. It learns the sequential motion from multiple stacks of slices and integrates the features between 2D slices and reconstructed 3D volume using affinity fusion, which resembles the iterations between slice-to-volume registration and volumetric reconstruction in the regular pipeline. The method accurately estimates the motion regardless of brain orientations and outperforms other state-of-the-art learning-based methods on the simulated motion-corrupted data, with a 48.4\% reduction of mean absolute error for rotation and 61.3\% for displacement. We then incorporated AFFIRM into the multi-resolution slice-to-volume registration and tested it on the real-world fetal MRI scans at different gestation stages. The results indicated that adding AFFIRM to the conventional pipeline improved the success rate of fetal brain super-resolution reconstruction from 77.2\% to 91.9\%.  
\end{abstract}

\begin{IEEEkeywords}
Motion correction, fetal magnetic resonance imaging, recursive framework, affinity fusion, slice-to-volume registration, super-resolution reconstruction.
\end{IEEEkeywords}

\section{Introduction}
\label{sec:introduction}
\IEEEPARstart{F}{etal} brain magnetic resonance imaging (MRI) is a powerful tool to assess fetal brain anomalies and dysfunction in prenatal examination [1]–[3], as it provides exquisite anatomical details and versatile contrasts compared to routine ultrasonography [4]. Direct 3D MRI of the fetal brain is desirable for quantitative analysis, It is, however, exceptionally challenging, as the image acquisition is highly susceptible to complex and irregular fetal and maternal movements. Instead, 2D multi-slice acquisition using single-shot fast spin-echo or balanced steady-state free precession is commonly used in the clinical setting, and the stacks of slices are acquired in multiple orthogonal views that are combined retrospectively to reconstruct the 3D volume of the fetal brain. Although 2D multi-slice acquisition freezes the in-plane motion, inter-slice motion is inevitable and imposes great challenges for 3D reconstruction. Thus, an accurate and robust motion correction framework is vital to improve fetal brain image analysis and facilitate applications in clinical practice.

Conventional retrospective motion correction methods involve slice-to-volume registration (SVR) and super-resolution reconstruction (SRR) of multiple 2D multi-slice images acquired in different orientations to solve an inverse problem [5]–[7]. In each iteration of the SVR-SRR pipeline, the low-resolution slices are registered to a target volume to correct the inter-slice motion and then construct into a high-resolution 3D volume using the putative motion parameters. Although the pipeline has already been implemented in several toolkits [8], [9], it sometimes fails due to the limited capture range and misalignment. Since the numerical optimization could not guarantee a globally optimal solution, SVR heavily relies on a relatively motion-free reference volume as well as good initializations of transformation parameters [10], which is difficult to estimate for large-scale motion. Moreover, neither single nor multi slice-to-volume registration methods take the sequential inter-slice motion patterns into account which encodes rich contextual information and probably improves the correction. 

Learning-based methods are recently emerging in fetal brain image analysis [11]–[13], including motion correction [14]. The convolution neural network (CNN) learns a data-driven function that maps the 2D slices to their corresponding locations in 3D space with high computational efficiency for fast motion tracking [15], [16]. The accuracy of the existing methods, however, may not be sufficient for motion correction of fetal brain MRI. One of the main reasons is the lack of an appropriate reference volume and the inference based on 2D slices is more analogous to pose approximation via image retrieval than structure-based methods [17]–[19]. Also, previous studies only fed 2D slices into the network and the motions of border slices are difficult to estimate given the insufficient information [15], [20]. As multiple stacks of slices with different views/orientations are readily available in fetal MRI examination, it is possible to choose or generate an appropriate reference and simultaneously integrate the features across stacks for better motion correction. The main challenge, nevertheless, lies in the dimensional correspondence between the features of the 2D slices and 3D volumes. Interestingly, the scheme of feature matching in unimodal image registration [21] can be considered as one special case of feature fusion, with two homogeneous sources with different spatial dimensions in this context. Inspired by the multi-modal fusion techniques in deep learning [22], [23], here, we present an attention-based feature affinity fusion method that learns the feature correspondence and explicitly builds a bridge between 2D slices and reference volumes, and thus, the network can also estimate the motion parameters in terms of their relation. 

\subsection{Related work}
3D Pose estimation, which localizes the subject pose in a single image, is a crucial problem in spatial reasoning [24]–[26]. The concept and relevant approaches, e.g., descriptor matching [27], [28] and absolute pose regression [29], [30], can be transferred to the problem of fetal motion estimation where the motion can be characterized by the continuous alternation of poses in the 3D canonical (atlas) space. Hou et al. used a CNN-based method, so-called SVRnet, to regress the rigid transformation parameters based on anchor-point slice parametrization and improved volumetric reconstructions for heavily moving fetuses [20]. Salehi et al. trained regression CNNs to estimate the 3D positions of the fetal brain using individual slices and also provided a fast SVR strategy [31]. Zhang et al. employed an end-to-end and multi-agent deep reinforcement learning network to detect the landmarks of the fetal pose in each time frame [32]. 

Despite the increased capture range and speed compared to conventional registration-based approaches, the above-mentioned studies treated the motion of each slice independently and the temporal information between slices, which are acquired in a sequential or interleaved fashion, is typically neglected. Since the motion trajectory of the fetal brain is continuous across time, strong temporal information can be assumed between adjacent slices [8], [33]. In light of this assumption, the slices can be considered as short-range dependent time series, and therefore, sequence learning-based methods, e.g., recurrent neural network (RNN), can be introduced in motion estimation. Singh et al. proposed a deep predictive motion tracking (DeepPMT) based on a long short-term memory network (LSTM) and achieved superior real-time fetal motion estimation performance compared to its alternatives [15]. Other advanced methods, such as gate recurrent unit (GRU) [34] and Vision Transformer (ViT) [35], which outperformed the LSTM in many tasks [36], may further improve the estimation performance. This work also adopted the idea to utilize the motion patterns by building bilateral connections between slices.

Well-designed multi-modal fusion schemes can improve the network learning capacity in many tasks. Among them, attention and gating mechanism are widely used to handle the features from multiple sources [37], [38]. Wang et al. developed a parallax-attention module to compute global correspondence in stereo images for super resolution [39]. Joze and Shaban et al. introduced multimodal excitation by gating mechanisms that allow fusion between different spatial dimensions at any level of the feature hierarchy [22]. You et al. proposed a novel cross-modality attention that uses semantic label embedding to guide integrations of spatial and temporal features [40]. Partly inspired by the cross attention mechanism that computes correlations across all possible features and captures the long-range dependency [41], [42], we adaptively modified it as a feature affinity fusion method to capture the global correspondence between the features extracted from the 2D slices and 3D volume.     
 
\subsection{Contributions}
To increase the success rate and accuracy for fetal brain motion correction, we proposed to join the merits of both the learning-based method and conventional SVR. The entire motion correction is therefore comprised of two steps as a coarse-to-fine procedure. We first introduced a novel Affinity Fusion-based Framework for Iteratively Random Motion (AFFIRM) correction, which replicated the iterative SVR and 3D volumetric reconstruction process, as a relatively coarse estimation of rigid motion parameters. The outcome of the AFFIRM was then used to initialize the multi-resolution SVR for fine-tuning.  

We first simulated motion-corrupted fetal brain T2-weighted MR images from 3D super-resolution reconstructed volumes, with gestational age (GA) from 23 to 40 weeks. Next, the proposed network was trained and tested on simulated motion-corrupted data and compared with other state-of-the-art methods. Lastly, we tested the AFFIRM-initialized pipeline with outlier rejection on the real-world fetal brain data that failed conventional SVR-SRR reconstruction.  

The contributions of this work are primarily three-fold. First, our model used bi-directional network architecture that takes the stack of slices as a time series to learn the sequential motion information and utilized complementary information across stacks. Second, the network employed a scattered data approximation (SDA) module for fast 3D reconstruction of the 2D stacks as well as an affinity fusion strategy to integrate the features between 2D slices and 3D volume. Third, we incorporated the proposed method into the conventional SVR as a coarse-to-fine process to tackle both small and large-scale motions and the pipeline increased the overall success rate of SRR. The AFFIRM demonstrated superior performance compared to the other existing learning-based fetal motion estimation methods on both simulated and real-world fetal brain MRI data.

The paper is organized as follows: the method and related theory are discussed in Section II. The experiments and the relevant results are described in Section III and IV, respectively, followed by a discussion in Section V and a conclusion in Section VI. The relevant code and demo have been released on Github (https://github.com/Allard-Shi/AFFIRM).

\section{Method}
\subsection{Problem formulation}
The fetal brain motion can be parameterized by 3D rigid transformation $\pmb{T}$ with 6 Degrees of Freedom (DoF). Among the various parameterizations for the transformation [25], we choose the Euler-Cartesian representation with three rotation parameters $\pmb{\theta} =(\theta_x,\theta_y,\theta_z)^T$ and three translation parameters $\pmb{d} =(d_x,d_y,d_z)^T$ in a 3D space. $\pmb{R} \in \mathbb{R}^{3\times 3}$ is the 3D rotation matrices based on $\pmb{\theta}$, and the rigid transformation $\pmb{T}$ can be represented as a 4 by 4 matrix:
\begin{equation}
\pmb{T}(\pmb{R},\pmb{d}) =\begin{bmatrix}
 \pmb{R} & \pmb{d} \\
  0&1
\end{bmatrix}
\label{eq}\end{equation}
The slices in each stack can be denoted as a temporal sequence $\left \{\pmb{X}\right \}_{n=1}^{N}$ where $N$ is the number of slices. Each slice is acquired given a pose of the fetal brain in the 3D standard space, and thus, corresponds to specific rigid motion parameters. All stacks were acquired in axial, coronal, and sagittal views of the fetal brain during the scan and are naturally divided into three groups in the following analysis [43]. The standard orientations are determined based on the three orthogonal orientations in the spatiotemporal fetal brain atlas [44]. 
\subsection{Affinity Fusion-based Framework for Iteratively Random Motion correction}
\subsubsection{Overall framework}
\begin{figure}[!t]
\centerline{\includegraphics[width=\columnwidth]{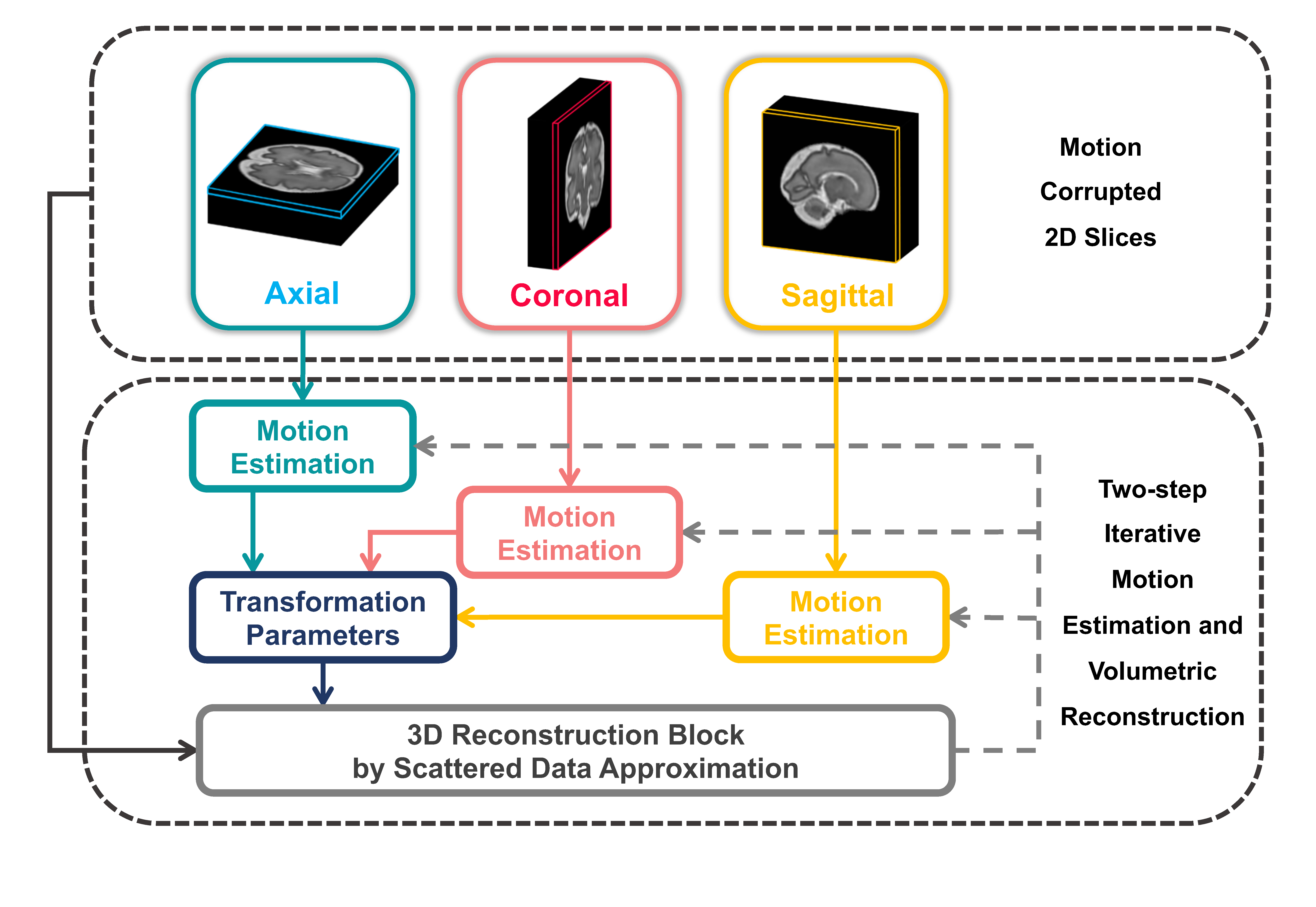}}
\caption{Schematic diagram of the AFFIRM. The model comprises motion estimation blocks for individual stacks and a 3D reconstruction block using scattered data approximation. Motion-corrupted 2D stacks from axial, coronal, and sagittal views in a range of oblique orientations are taken as inputs. The final estimated motion parameters are obtained after iterative updates between the motion estimation blocks and the reconstruction block.}
\label{fig1}
\end{figure}
The entire AFFIRM is illustrated in Fig 1. It is comprised of two types of blocks, i.e., motion estimation block (see Section B2) and 3D reconstruction block, which replicates the iterations between SVR and SRR. The network takes three motion-corrupted stacks of slices in axial, coronal, and sagittal orientations as inputs into learning-based motion estimation blocks to predict a set of motion parameters. A 3D volume is then roughly reconstructed using SDA [8], [45] based on multiple stacks and the estimated motion parameters in the current iteration. Briefly in the SDA, the scattered data, i.e., multiple stacks of 2D slices, is sampled onto a 3D regular grid based on the nearest neighbor scheme, followed by a Gaussian blurring operation. The blurring kernel size defined in the standard deviation monotonically decreases from 0.8 mm for the first iteration to 0.52 mm for the last iteration given an approximated point spread function of the acquisition model [6], [33], [46]. Hence, the reconstruction procedure, albeit coarse, rapidly provides a volumetric fetal brain reference in the standard anatomical space. Next, the volume together with the original stacks of slices are again fed into the motion estimation block and reconstruction module to update the motion parameters and refine the reference volume. The iteration ends when a predefined number of iterations is reached. Note that we manually selected the stack with relatively small motion and resampled and rigidly aligned it to the fetal brain atlas [44] using FLIRT [47] as the initial reference volume. AFFIRM is trained by an end-to-end scheme and the network parameters at each iteration are shared and updated together.   

\subsubsection{Motion estimation block}
The motion estimation block of AFFIRM is illustrated in Fig 2a, which predicts the rigid motion parameters $\pmb{T}$ for each slice. A lightweight 2D CNN is first applied to all slices to generally extract informative features, and simultaneously, a reference volume is fed into a 3D CNN. The two convolution-based networks are composed of stacks of standard CNN units, i.e., a convolutional layer followed by a Batch Normalization unit, a Rectified Linear Unit (ReLU), and a pooling operation, with detailed architectures shown in Fig 2b. 

\begin{figure*}[!t]
\centerline{\includegraphics[width=\textwidth]{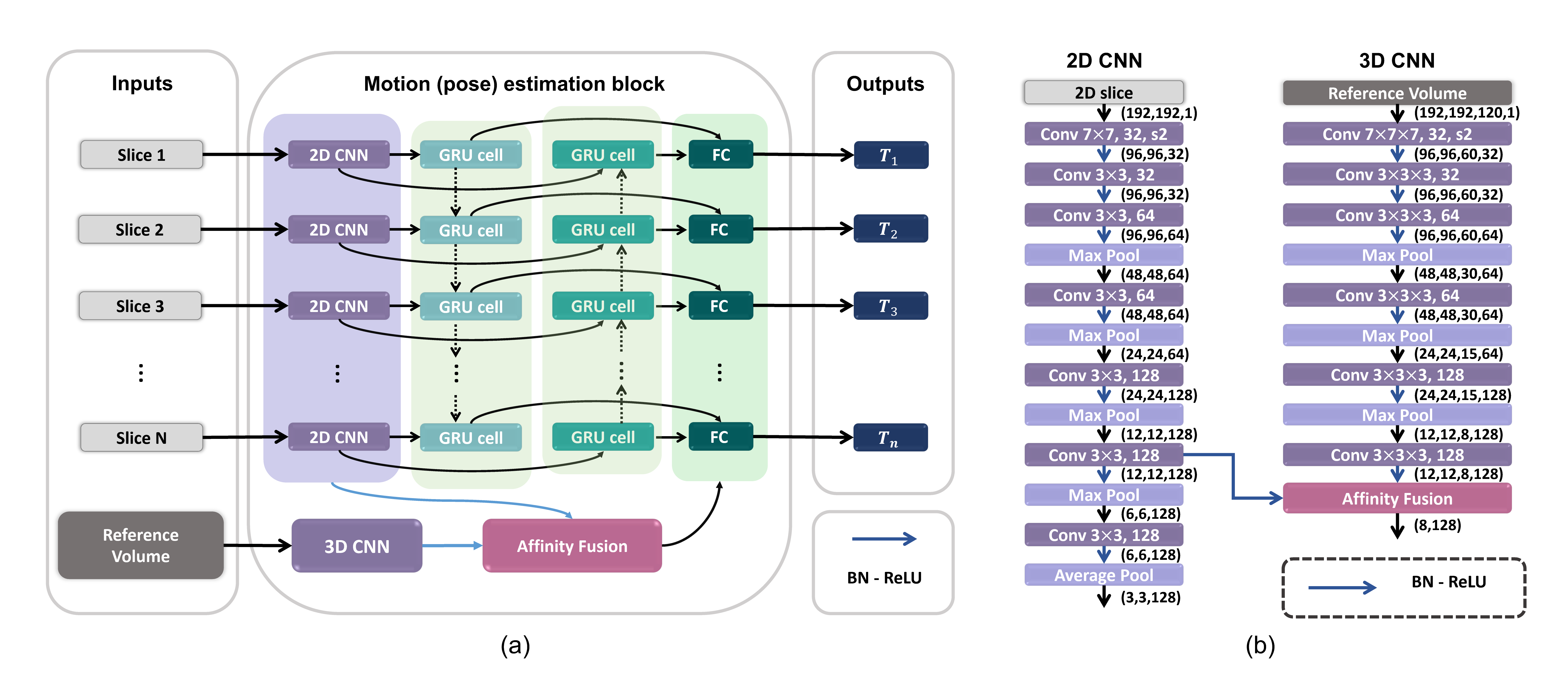}}
\caption{(a) Schematics of the motion correction block. (b) Architectures of the 2D CNN and 3D CNN with affinity fusion. FC $=$ fully connected layer. BN $=$ batch normalization.}
\label{fig2}
\end{figure*}

The 2D feature sequence then goes through a bidirectional recurrent network which learns the sequential motion patterns across slices. We choose the GRU, a computationally efficient variant of LSTM, as the backbone of the network, which includes the gating mechanism and learns the long-term dependency considering possible outliers. Each GRU cell accepts the information from the previous and current states, and the gates control the amount of information to be released using the sigmoid function. The bidirectional architecture is designed to enhance the motion pattern between successive slices. A fully gated GRU cell is implemented as:
\begin{equation}
\pmb{z}_i=\sigma_g(\pmb{W}_{zx}\pmb{x}_i+\pmb{W}_{zh}\pmb{h}_{i-1}+\pmb{b}_z)
\label{eq}\end{equation}
\begin{equation}
\pmb{r}_i=\sigma_g(\pmb{W}_{rx}\pmb{x}_i+\pmb{W}_{rh}\pmb{h}_{i-1}+\pmb{b}_r)
\label{eq}\end{equation}
\begin{equation}
\pmb{h}_i=(1-\pmb{z}_i)\odot \pmb{h}_{i-1}+\pmb{z}_i\odot\sigma_h(\pmb{W}_{hx}\pmb{x}_i+\pmb{W}_{hr}(\pmb{r}_i\odot\pmb{h}_{i-1})+\pmb{b}_h)
\label{eq}\end{equation}
where $\pmb{x}_i$, $\pmb{z}_i$, $\pmb{r}_i$, and $\pmb{h}_i$ denote the input of GRU, i.e., 2D features from CNN, the update gate, reset gate, and the hidden state for the time frame $i$ in the sequence, respectively. $\pmb{W}_{zx}$ and $\pmb{W}_{zh}$ are weights associated with update gate while $\pmb{W}_{rx}$ and $\pmb{W}_{rh}$ are with reset gate. $\pmb{W}_{hx}$ and $\pmb{W}_{hr}$ are weights that determine the output. $\pmb{b}_z$, $\pmb{b}_r$ and $\pmb{b}_h$ are bias terms. $\sigma_g\cdot$ is the sigmoid function and $\sigma_h\cdot$ is the hyperbolic tangent function tanh. $\odot$ represents the Hadamard product. The output of each GRU cell $\pmb{h}_i$ is then passed into two heads, and each has a fully connected layer with ReLU activation function and incorporates the Dropout [48]. Lastly, a range-dependent $tanh$ activation is added on top of the layer to predict the transform parameters for each time step. 

Inspired by the cross-attention that is a special case of non-local operations [41], we proposed an affinity fusion method to exploit and utilize the relationship between the 2D and 3D feature maps for estimating $\pmb{T}$. Specifically, the intermediate feature maps $\pmb{x}_s \in \mathbb{R}^{H\times W\times C}$ in the 2D CNN and $\pmb{x}_v \in \mathbb{R}^{H\times W\times D\times C}$ in the 3D CNN are flattened, where $H$, $W$, $D$ denote the shapes of the features and $C$ is the number of channels. The output response $\pmb{y}_j$, namely the fusion operation, at the position $j$ can be generally formulated as:
\begin{equation}
\pmb{y}_j=\frac{1}{h(\pmb{x}_s,\pmb{x}_v)}  {\textstyle \sum_{k}} a(\pmb{x}_s^j,\pmb{x}_v^k)g(\pmb{x}_v^k)
\label{eq}\end{equation}
where  $a(\cdot)$  is a function that computes the relationship, i.e., affinity, between 2D feature maps $\pmb{x}_s$ at spatial position $j$ and 3D feature maps $\pmb{x}_v$ at spatial position $k$. The function $g$ calculates a representation of 3D feature maps $\pmb{x}_v$ at spatial position $k$, and usually, a simple linear embedding is adequate to generalize an effective representation, i.e., $g(\pmb{x}_v^k)=\pmb{W}_g\pmb{x}_v^k$ where $\pmb{W}_g$ is a weighted matrix learned by the network. To facilitate learning, the response $\pmb{y}_j$ is commonly normalized by a factor $h(\cdot)$. In this study, the affinity function $a(\cdot)$ is defined as a Gaussian function in the embedding space, so the fusion output can be rewritten as:
\begin{equation}
\pmb{y}=\sigma_s(\frac{(\pmb{W}_s\pmb{x}_s)^T\pmb{W}_v\pmb{x}_v}{\sqrt{C}})g(\pmb{x}_v)
\label{eq}\end{equation}
where $\sigma_s(\cdot)$ is the softmax function. $\pmb{W}_s$ and $\pmb{W}_v$ are weighted matrices to be learned. The operation, namely, the scaled dot product attention [37], encodes the spatial relationship between the slices and volume. The affinity fusion is applied for 2D feature maps $\pmb{x}_s$ for each slice separately and the response $\pmb{y}$ for a given slice is subsequently concatenated to the output of the bidirectional network of the same slice.    

\subsubsection{Loss function}
Loss of the proposed network is composed of two parts: 1) regression loss of the transformation parameters ($\mathcal{L}_1$) and 2) consistency loss of the motion-corrected volume ($\mathcal{L}_2$). For the regression loss, since the set of 3D rigid motion is a special Euclidian group denoted as $SE(3)$, we used the geodesic distance as the loss to measure the distance on the 6-dimensional non-Euclidean manifold [31] instead of the Euclidean norms. We designed the loss $\mathcal{L}_1$ for the transform parameters $\pmb{T}$ as: 
\begin{equation}
\mathcal{L}_1(\pmb{T},\hat{\pmb{T}})=\sum_{s\in S} ({\left \| log(\hat{\pmb{R}_s}^T\pmb{R}_s) \right \|}_F^{2}  +\gamma\left \|\hat{\pmb{d}_s}- \pmb{d}_s \right \|_F^2)^{\frac{1}{2}}
\label{eq}\end{equation}
where $\pmb{R}_s$ and $\pmb{d}_s$ are the rotation matrix and displacement parameters of slice $s$, respectively. $\left \| \cdot  \right \| _F$ is the Frobenius norm and $log(\cdot)$ represents the matrix logarithm. $S$ denotes the 2D slices containing the fetal brain, and $\gamma$ is a hyperparameter to balance the rotation and displacement units. 

We defined loss of reconstructed volume $\mathcal{L}_2$ as:
\begin{equation}
\mathcal{L}_2(\pmb{I},\pmb{V},\hat{\pmb{T}})=\left \| \Omega (\pmb{I},\hat{\pmb{T}})-\pmb{V}\right \|_2
\label{eq}\end{equation}
in which $\pmb{I}$ is the available stacks of 2D slices, and $\pmb{V}$ is the motion-free fetal brain volumes. $\hat{\pmb{T}}$ is the final estimated motion parameters and $\Omega(\cdot)$ represents the SDA. $\left \|  \cdot\right \| _2$ denotes the $L_2$ norm. 

The final loss function of AFFIRM is defined as a weighted combination of $\mathcal{L}_1$ and $\mathcal{L}_2$, where $\lambda$ is a weighting hyperparameter. 
\begin{equation}
\mathcal{L}=\mathcal{L}_1(\pmb{T},\hat{\pmb{T}})+\lambda\mathcal{L}_2(\pmb{I},\pmb{V},\hat{\pmb{T}})
\label{eq}\end{equation}

\section{Experiment}
\subsection{Data acquisition and preprocessing}
In-utero MRI of 180 fetal brains acquired in at least three orthogonal orientations was collected under Institutional Review Board approval. The data were acquired on a 3T Siemens Skyra scanner (Siemens Healthineers, Erlangen, Germany) with an abdominal coil, using a T2-weighted half-Fourier single-shot turbo spin-echo (HASTE) sequence with the following protocol: TR/TE = 800/113 ms, in-plane resolution = 1.09 × 1.09 mm, matrix = 256 × 200, 35-55 slices with slice thickness = 2 mm, partial Fourier factor = 5/8, echo spacing = 5.36 ms, echo train length = 102, GRAPPA factor = 2, number of calibration line = 42 with an interleaved acquisition. Altogether 31 cases were excluded due to low image quality, such as low signal-to-noise ratio (SNR) and severe signal voids. 

The data preprocessing conforms to the following procedure. First, the bias field was corrected using the N4 algorithm [49]. Then, inter-plane motion correction and SRR at a 0.8 mm isotropic spacing were performed via the NiftyMIC toolkit\footnote{https://github.com/gift-surg/NiftyMIC} [8] . The fetal brain was extracted by manual masking using ROIEditor\footnote{https://www.mristudio.org/}, followed by 3D non-local means denoising [50]. Each stack was normalized to the maximal intensity. Next, the reconstructed brain volumes were rigidly registered to the spatiotemporal fetal brain atlas in [44] using FLIRT [47], center-aligned, and resized into 192×192×144 by padding zeros in the surrounding regions. Altogether, 34 cases failed in the SRR step due to motion-induced misalignment, and the remaining 115 subjects (GA: 23.1-40.0 weeks; mean=30.62 weeks, standard deviation=4.41 weeks) were finally used in the current study. Note that the super-resolution reconstructed volumes from the 2 mm-thick slices were used as the ground truth.

\subsection{Motion simulation}
To simulate the dynamic fetal brain motion, we adopted a control-point-based scheme [15] that generates a continuous motion trajectory given several control points based on a simple random walk model [51]. Specifically, for one motion parameter, $N_c$ control points $\left \{P\right \}_{k=1}^{N_c}$  are generated by $P_k=P_{k-1}+\Delta P$, where $\Delta P$ and $P_1$ are drawn from specific uniform distributions. Univariate smoothing cubic splines are then used to fit a curve, i.e., the motion trajectory, based on the control points and the curve is subsequently demeaned. Next, a random offset is added to control the mean of the continuous curve, and the motion parameter for each slice can be uniformly sampled from the curve (see Fig. 3b). The sets of control points correspond to various motion patterns and thus generate infinite stacks of slices as training sequences based on the acquisition model. We only randomly sampled each motion parameter rather than the 6D manifold as these sampling strategies had little influence on the prediction accuracy given adequate training data [20].    

\begin{figure}[!t]
\centerline{\includegraphics[width=\columnwidth]{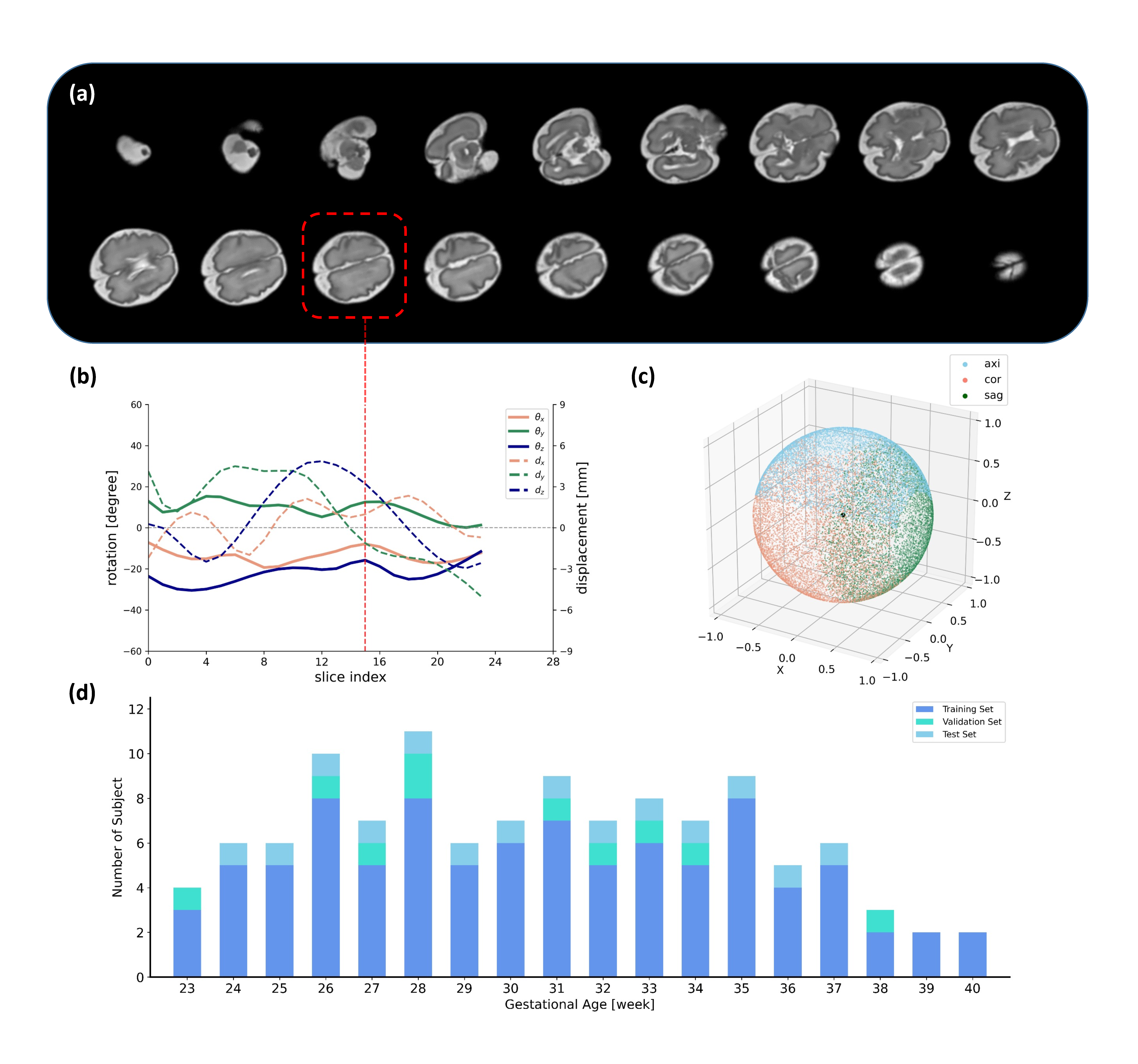}}
\caption{Simulation of the inter-slice motion. (a) Simulated motion-corrupted slices of a 29-week GA fetal brain (slice thickness=4mm) acquired in the axial orientation. (b) The simulated trajectory of the motion parameters. (c) Fetal brain orientations in 3D space characterized by the slice plane normal with respect to the origin (axi=axial; cor=coronal; sag=sagittal) over a unit sphere. (d) Gestational age distributions of the training (N=91), validation (N=10), and test set (N=14).}
\label{fig3}
\end{figure}

In this work, the mean of the simulated Euler angles $\bar{\pmb{\theta}}$, controlled by the random offset, was bound to a uniform distribution $U$ in the range between $-\frac{\pi}{4} $ and $\frac{\pi}{4} $, such that $\bar{\pmb{\theta}}\sim U(-\frac{\pi}{4},\frac{\pi}{4})$, to avoid the gimbal lock phenomenon. The mean angular velocity was set below 5 degrees per second for each rotation axis. The corresponding transform matrix is also adapted to the coronal and sagittal stacks. The simulation, thus, covers all possible fetal brain orientations in the 3D space (Fig. 3c). For translational motion, as the stacks are center-aligned beforehand, we controlled $\pmb{d}\in (-10,10)$ and $\bar{{\pmb{d}} } \in (-2,2)$ in millimeter. The motion-corrupted data were finally generated according to the imaging acquisition model and motion parameters [8]. As the real multi-slice 2 mm-thick data were acquired in an interleaved order ([1, 3, 5, …, 2, 4, 6, …]), every other slice with a 4 mm gap made a continuous motion. Therefore, for simplicity, we performed simulation on 4 mm-thick slices drawn from the 3D volumes (non-interleaved, no gap) that typically consists of 24 slices covering the whole brain.

\subsection{Algorithm comparison}
We compared AFFIRM with other existing state-of-the-art learning-based motion estimation approaches in fetal MR images, including anchor-point-based SVRnet [16], [20], Deep Pose Estimation (DeepPE) based on an 18-layer residual network [31], and DeepPMT proposed in [15]. For a fair comparison, the estimated anchor point labels in SVRnet were transformed to Euler-Cartesian representation. VGGNet was used as the skeleton of SVRnet, which was reported to attain the least regression errors. Since DeepPE and DeepPMT could not take the in-plane displacement into account, they were not compared regarding translational motion. All the networks were trained using the same motion simulation. The prediction error of the motion parameters was only compared for the axial acquisition as DeepPMT cannot handle the slices acquired in the coronal and sagittal orientations. Multiresolution conventional SVR was also included in the comparison. Without loss of generality, we performed the SRR based on the simulated motion-corrupted stack of 2D slices in the axial orientation and 2 motion-free stacks in the coronal and sagittal orientations. The reconstruction was formulated as a convex problem in a maximum-a-posteriori (MAP) manner and solved using a dedicated linear least-squares solver, as implemented in the NiftyMIC [8]. All the reconstructed volumes were rigidly aligned to the corresponding ground truth for quantitative evaluation.

Mean Absolute error (MAE) and root mean square error (RMSE) were used as the evaluation metrics to measure the prediction accuracy of motion parameters. Only the slices including the brain tissues are taken into consideration. Structural similarity (SSIM) was employed to compare the similarity between the reconstructed volumes and the ground truth motion-free volumes. Structural dissimilarity (DSSIM) maps derived from SSIM were calculated to visualize the regional dissimilarity [52]. Normalized root mean square error (NRMSE) ranging between 0 and 1 was used as a complementary metric for assessing the reconstruction quality against ground truth.

\subsection{Ablation experiment}
We performed ablation studies to evaluate the individual components of AFFIRM. To assess the fusion methods and the effects of network recurrence on the motion estimation accuracy, we trained four networks:
1)	A non-recursive network without fusion that only retained the 2D branch, bidirectional network, and the subsequent FC module.
2)	A non-recursive network with late 2D/3D fusion. Specifically, the late fusion approach only fed the 3D branch into the fully connected layer by a simple concatenation operation, which is commonly used in multi-modality learning [53]–[55]. 
3)	A non-recursive network with 2D/3D affinity fusion proposed in this work. 
4)	A recursive network with affinity fusion, i.e., AFFIRM.
All these networks were trained under the same schemes and tested on the same simulated motion-corrupted data for comparisons. We performed paired t-tests between schemes on the MAE and RMSE of rotation and displacement, followed by Bonferroni correction for multiple comparisons. All the statistical tests in this work are performed using the Python SciPy statistical library.  

\subsection{Coarse-to-fine motion correction}
The toolkit NiftyMIC implements principal-brain-axes-based initialization in the volume-to-volume registration based on normalized cross correlation (NCC) and initializes the SVR using the center of geometry [8]. The registration process is relatively robust when the fetal motion is small. However, we found some subjects (N=34 out of 149) failed the NiftyMIC pipeline due to large motion-induced misalignment or inaccurate initialization. Here, we integrated AFFIRM and the conventional multi-resolution SVR, which serves as a coarse-to-fine motion correction scheme. It was then combined with the SRR into an automated two-step framework with slice rejection. We then used the AFFIRM-SVR pipeline for motion correction and 3D brain reconstruction on the test data and those that failed NiftyMIC. The real-world image slices were collected in an interleaved fashion and were fed into our approach by dividing 2 mm-thick stacks into two sets of continuous series.

\subsection{Implementation details}
The set of 115 fetuses was split into 91, 10, and 14 subjects as training, validation, and testing subjects, respectively. The GA range of the training set spans over 23.1 to 40.0 weeks to enhance the network generalizability over large anatomical variations across gestation. The specific GA distributions of the training, validation, and test sets are shown in Fig. 3d.  

The network was implemented using TensorFlow and Keras library and trained on four NVIDIA GeForce RTX 2080Ti GPUs. A recurrence number of four was empirically adopted considering the computational efficiency and a two-step training strategy was used. First, we used the mean square error (MSE) loss of the transform parameters, i.e., (5) in [31], and the RMSprop optimizer in the initial training session, with a learning rate of  $10^{-4}$, and reduced the learning rate by a factor 2 if there was no improvement on the validation set for 4 epochs. The network was then trained using the loss defined in (9) when the learning rate was below $10^{-5}$. In each epoch, the network was trained on random 150 sets of simulated motion-corrupted stacks of slices in the axial, coronal, and sagittal orientations respectively, using the scheme in Section III.B, and the training session stopped after 100 epochs. Data augmentation was performed by varying the fetal brain volume from 0.875 to 1.125 times the original size in the even step, resulting in an augmentation fold of 10 per subject. 

\section{Result}
The comparison of fetal motion estimation algorithms is shown in Table I in terms of rotation and displacement. AFFIRM outperformed the other state-of-the-art learning-based methods with the lowest average MAE of 4.83 degrees for rotation and 1.52 mm for displacement, and the lowest average RMSE of 6.07 degrees for rotation and 1.91 mm for displacement, which was more than 48.4\% reduction of MAE for rotation parameters and 61.3\% for displacement in compared to the other methods. Fig. 4 shows exemplary results of estimated fetal brain motion at different GA and orientations in the test set using AFFIRM, which showed high consistency with the ground truth. Fig. 5 shows the estimation accuracy of AFFIRM over the GA from 24 weeks to 37 weeks in terms of MAE and RMSE for rotation and displacement. The Spearman's rank correlation showed that there was obvious GA-dependency for both MAE of rotation ($\rho$=-0.227, $p$=0.007) and displacement parameters ($\rho$=-0.309, $p$=0.002) and RMSE of rotation ($\rho$=0.377, $p$\textless$10^{-5}$) and displacement parameters ($\rho$=0.234, $p$=0.002).

\begin{table}
\caption{Comparison of different learning-based methods of fetal motion estimation. The estimation errors for rotation and displacement motion in the axial stacks are listed}
\label{table}
\begin{tabular}{ccccc}
\hline
\toprule[0.75pt]
\multirow{2}{*}{Method} & \multicolumn{2}{c}{Rotation (degree)} & \multicolumn{2}{c}{Displacement (mm)} \\ \cline{2-5} 
                                       & MAE         & RMSE           & MAE           & RMSE            \\ \hline
DeepPE [31]                            & 21.29±6.46  &24.17±6.47      & -             &  -              \\
SVRnet [20]                            & 10.63±3.27  &14.16±3.96      &3.93±0.92      & 7.23±1.87       \\
DeepPMT [15]                           & 9.35±3.02   &12.06±3.48      & -             & -               \\
AFFIRM (Ours)                          & \textbf{4.83±1.27}   & \textbf{6.07±1.58}       &\textbf{1.52±0.34}      & \textbf{1.91±0.40}       \\ 
\bottomrule[0.75pt]
\hline

\multicolumn{5}{p{251pt}}{MAE $=$ mean square error; RMSE $=$ root mean square error; DeepPE $=$ deep prediction estimation; DeepPMT $=$ deep predictive motion tracking; AFFIRM $=$ affinity fusion-based framework for iteratively random motion correction.}

\end{tabular}
\end{table}

\begin{figure*}[!t]
\centerline{\includegraphics[width=\textwidth]{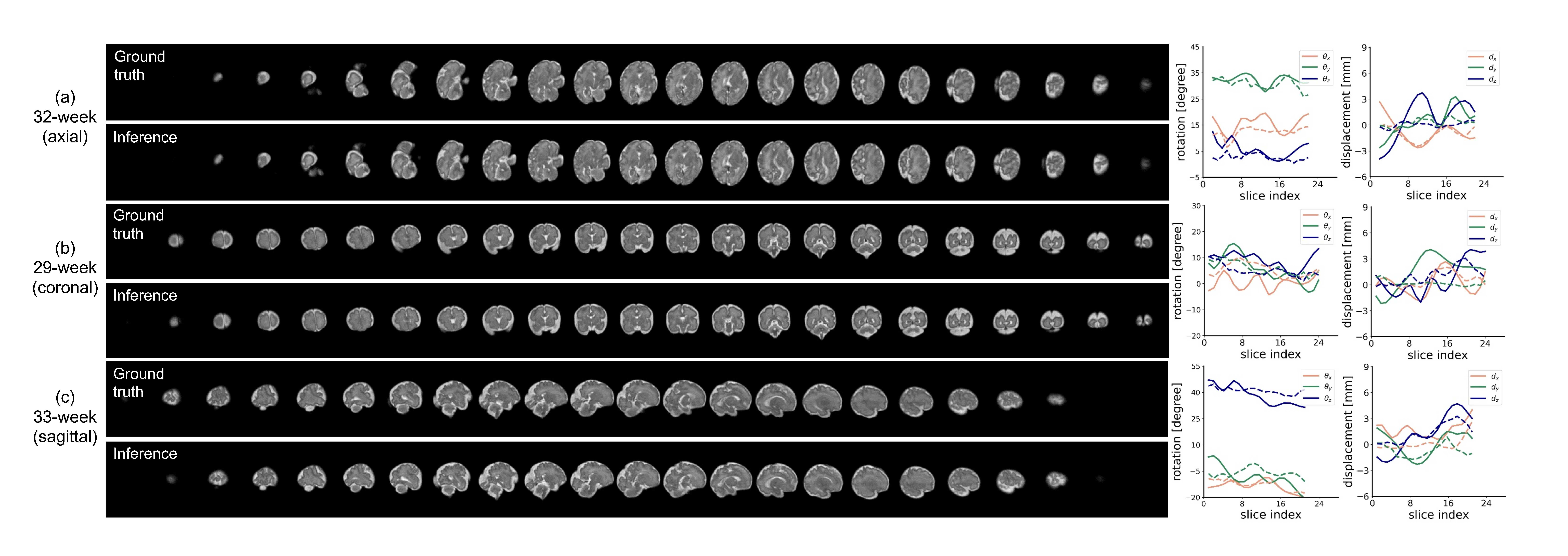}}
\caption{Estimated fetal brain motion (inference) using AFFIRM, compared with the simulated motion-corrupted data (ground truth). Simulated scans of three fetuses at (a) 32-week GA in axial view, (b) 29-week GA in coronal view, and (c) 33-week GA in sagittal view, with relevant motion trace on the right (solid curve: ground truth, dotted curve: inference). }
\label{fig1}
\end{figure*}

\begin{figure}[!t]
\centerline{\includegraphics[width=\columnwidth]{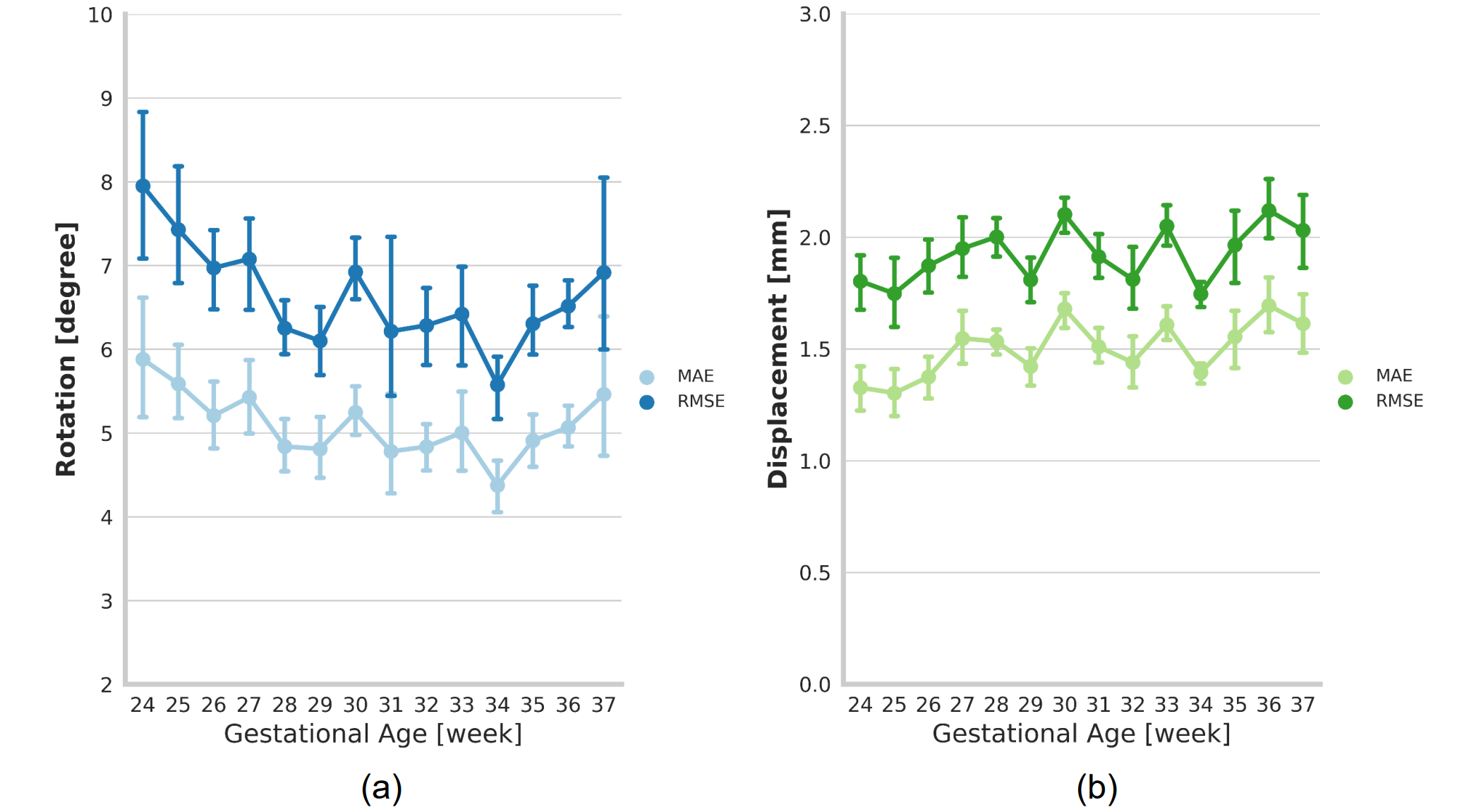}}
\caption{Effect of gestational age on the estimation errors of (a) rotational and (b) displacement fetal brain motion using AFFIRM. MAE $=$ mean absolute error; RMSE $=$ root mean square error.}
\label{fig1}
\end{figure}

Fig. 6 shows that super resolution-reconstructed images using AFFIRM-corrected data achieved the highest SSIM of 0.961 and the lowest NRMSE of 0.211 among all learning-based and the multiresolution SVR method on the simulated motion-corrupted data. The final SRR results in Fig. 7 reveal that better volumetric reconstruction is achieved using AFFIRM, e.g., lower DSSIM can be visually appreciated.

\begin{figure}[!t]
\centerline{\includegraphics[width=\columnwidth]{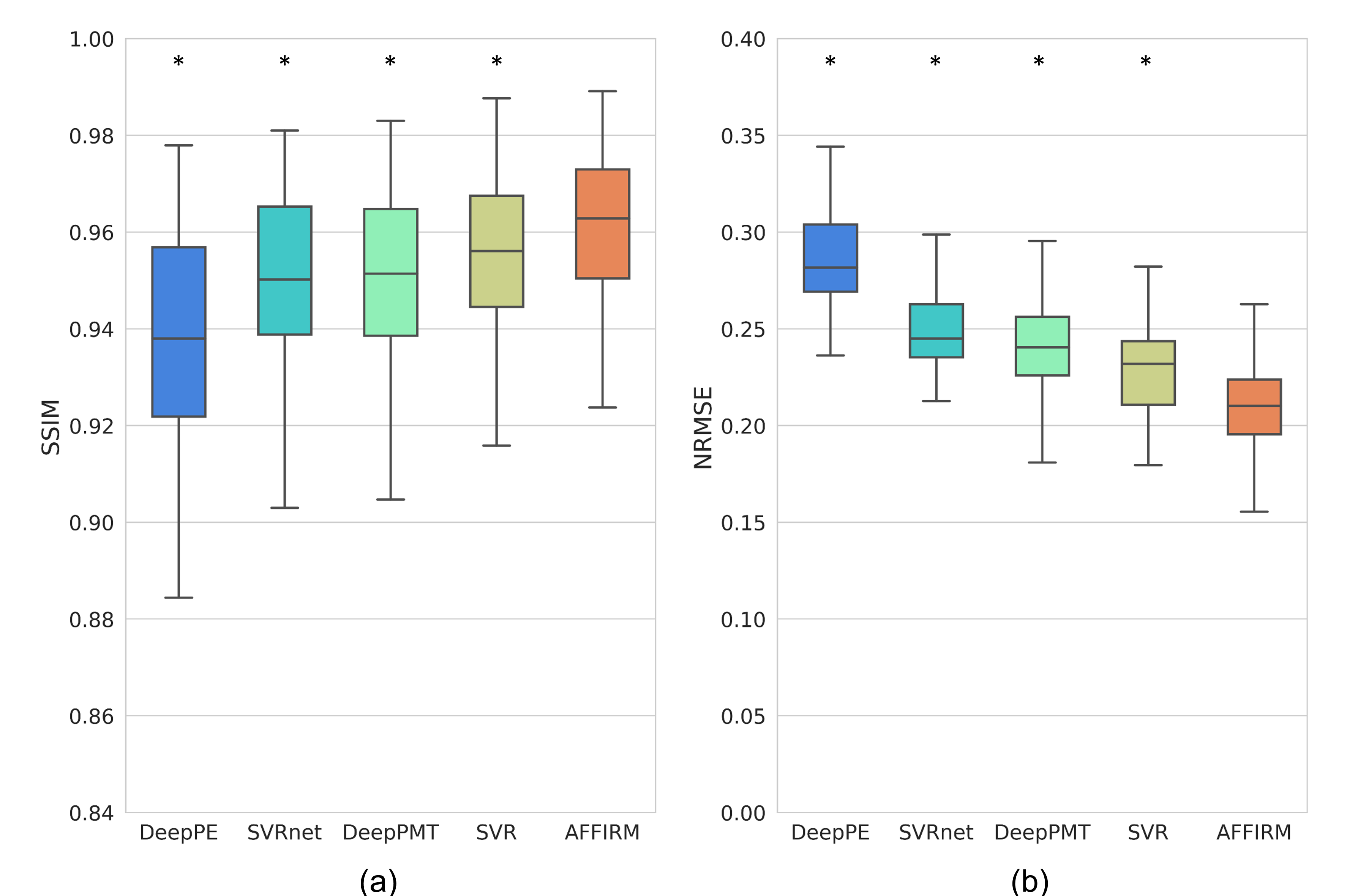}}
\caption{Comparison of different motion correction methods in terms of super-resolution reconstruction. The asterisk labels the significant difference ($p$\textless$10^{-10}$) with respect to AFFIRM using paired t-test with Bonferroni correction. DeepPE $=$ deep prediction estimation; DeepPMT $=$ deep predictive motion tracking; SVR $=$ slice-to-volume registration; AFFIRM $=$ affinity fusion-based framework for iteratively random motion correction; SSIM $=$ Structural similarity; NRMSE $=$ normalized root mean square error.  }
\label{fig6}
\end{figure}

\begin{figure*}[!t]
\centerline{\includegraphics[width=\textwidth]{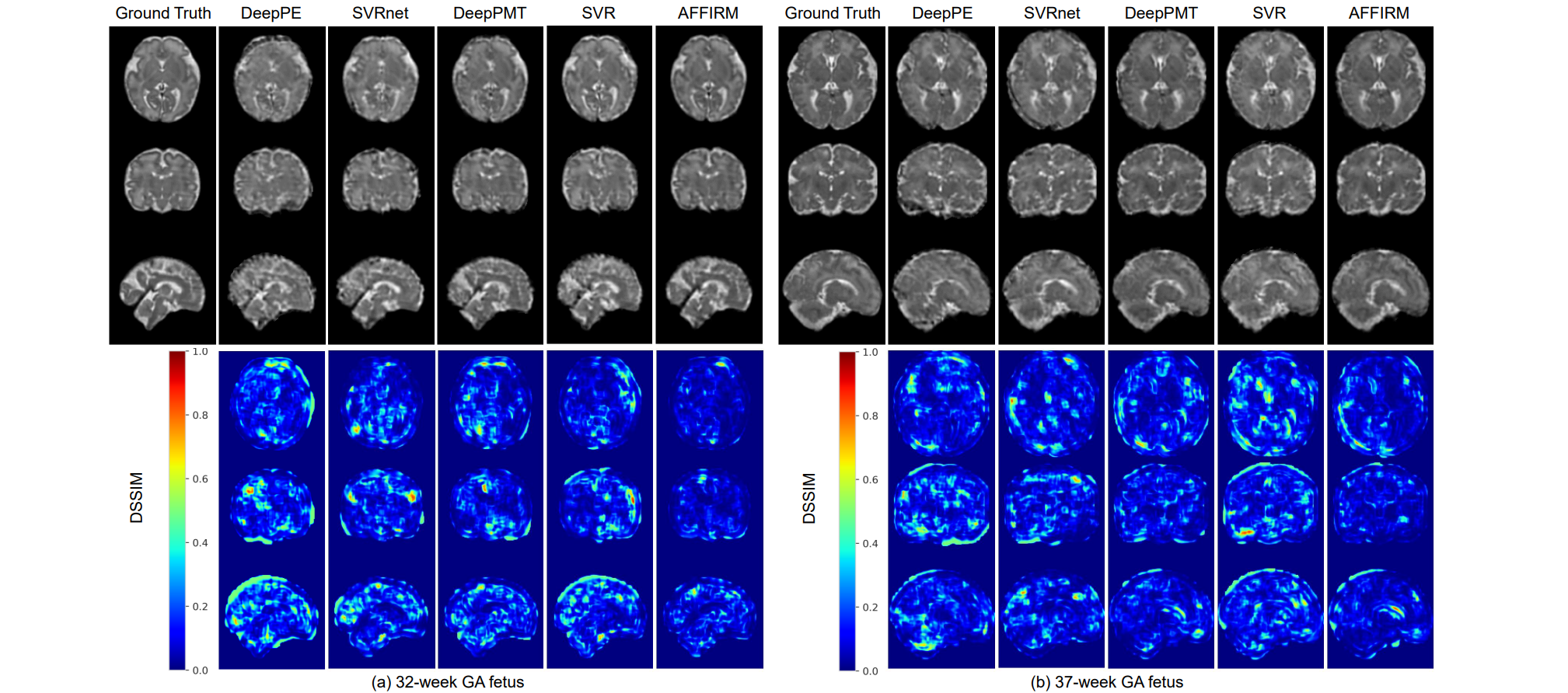}}
\caption{Super-resolution reconstruction of the simulated motion-corrupted data for representative fetal brains at (a) 32-week GA and (b) 37-week GA using DeepPE, SVRnet, DeepPMT, multi-resolution hierarchical SVR, and AFFIRM schemes in comparison with the ground truth. DSSIM $=$ structural dissimilarity.}
\label{fig7}
\end{figure*}

We further evaluated the effects of fusion and recurrence on the network performance with ablation experiments. The statistical testing results in Table II showed no significant reduction of MAE and RMSE in rotation and even a significant increase of MAE ($t$=-4.682, $p$\textless$10^{-4}$) and RMSE ($t$=-4.543, $p$\textless$10^{-4}$) in displacement using the network with late fusion compared with that without fusion. There are significant reductions of MAE and RMSE in rotation ($t_{MAE}$=5.827, $p_{MAE}$\textless$10^{-6}$; $t_{RMSE}$=6.090, $p_{RMSE}$\textless$10^{-7}$) and displacement ($t_{MAE}$=5.022, $p_{MAE}$\textless$10^{-5}$; $t_{RMSE}$=2.744, $p_{RMSE}$=0.041) using the network with 2D/3D affinity fusion compared to that without fusion. Furthermore, we also found significant reduction of MAE and RMSE in rotation ($t_{MAE}$=6.321, $p_{MAE}$\textless$10^{-7}$; $t_{RMSE}$=5.989, $p_{RMSE}$\textless$10^{-6}$) and displacement ($t_{MAE}$=7.544, $p_{MAE}$\textless$10^{-10}$; $t_{RMSE}$=6.367, $p_{RMSE}$\textless$10^{-7}$) using AFFIRM compared to the non-recursive network with 2D/3D affinity fusion. 

\begin{table}
\caption{The ablation experiment results for evaluating the effects of the affinity fusion and network recurrence on the rotation and displacement motion.}
\label{table}
\begin{tabular}{ccccc}
\hline
\toprule[0.75pt]
\multirow{2}{*}{Network architecture} & \multicolumn{2}{c}{Rotation (degree)} & \multicolumn{2}{c}{Displacement (mm)} \\ \cline{2-5} 
                                              & MAE          & RMSE             & MAE           & RMSE        \\ \hline
No fusion   & \multirow{2}{*}{5.59±1.01}   & \multirow{2}{*}{7.32±1.38}                                                                            & \multirow{2}{*}{1.59±0.21}   & \multirow{2}{*}{2.00±0.23}    \\
non-recursive\\

Late fusion   & \multirow{2}{*}{5.76±1.16}   & \multirow{2}{*}{7.47±1.56}                                                                            & \multirow{2}{*}{1.61±0.21}   & \multirow{2}{*}{2.03±0.23}    \\
non-recursive \\

Affinity fusion   & \multirow{2}{*}{5.29±0.88}   & \multirow{2}{*}{6.83±1.21}                                                                            & \multirow{2}{*}{1.54±0.20}   & \multirow{2}{*}{1.97±0.23}    \\
non-recursive\\

Affinity fusion   & \multirow{2}{*}{\textbf{5.10±0.84}}  & \multirow{2}{*}{\textbf{6.64±1.15}}                                                                            & \multirow{2}{*}{\textbf{1.50±0.20}}   & \multirow{2}{*}{\textbf{1.92±0.23}}    \\
recursive (AFFIRM) \\

\bottomrule[0.75pt]
\hline

\multicolumn{5}{p{251pt}}{no fusion $=$ network with only a 2D branch. late fusion $=$ non-recursive network with late concatenation fusion between 2D and 3D branches. affinity fusion $=$ non-recursive network with affinity fusion between 2D and 3D branches. AFFIRM $=$ affinity fusion-based framework for iteratively random motion correction (recursive network with affinity fusion between 2D and 3D branches); MAE $=$ mean absolute error; RMSE $=$ root mean square error.}

\end{tabular}
\end{table}

Lastly, as described in Section III.E, we plugged AFFIRM as a novel initialization step in the multi-resolution SVR process and tested the entire pipeline on the real-world fetal MRI scans that failed the conventional pipeline. The flow chart in Fig. 8 summarizes the volumetric reconstruction results. Specifically, NiftyMIC successfully reconstructed 115 subjects out of a total of 149 subjects, leading to a success rate of 77.2\%. The AFFIRM-SVR scheme further rescued 22 fetal brains out of the 34 failed subjects, and thus increased the success rate to 91.9\%. Note that all data in the test set (N=14) were successfully reconstructed by NiftyMIC with or without AFFIRM. Fig. 9 showed the reconstruction volumes from the AFFIRM-SVR scheme in comparison with the results that initially failed the conventional pipeline NiftyMIC.

\begin{figure}[!t]
\centerline{\includegraphics[width=\columnwidth]{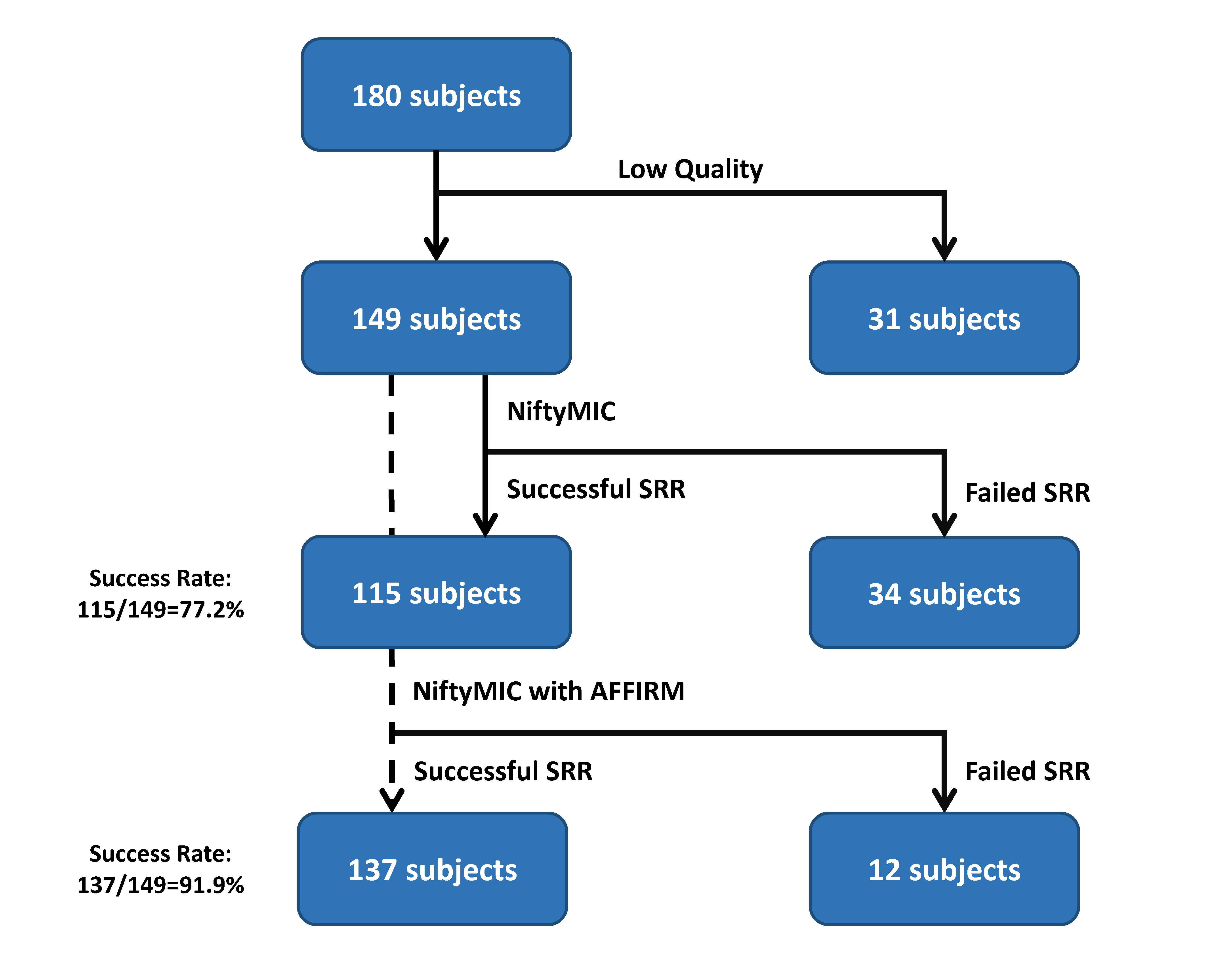}}
\caption{Success rate of fetal brain super-resolution reconstruction using NiftyMIC with or without the AFFIRM scheme.}
\label{fig1}
\end{figure}

\begin{figure}[!t]
\centerline{\includegraphics[width=\columnwidth]{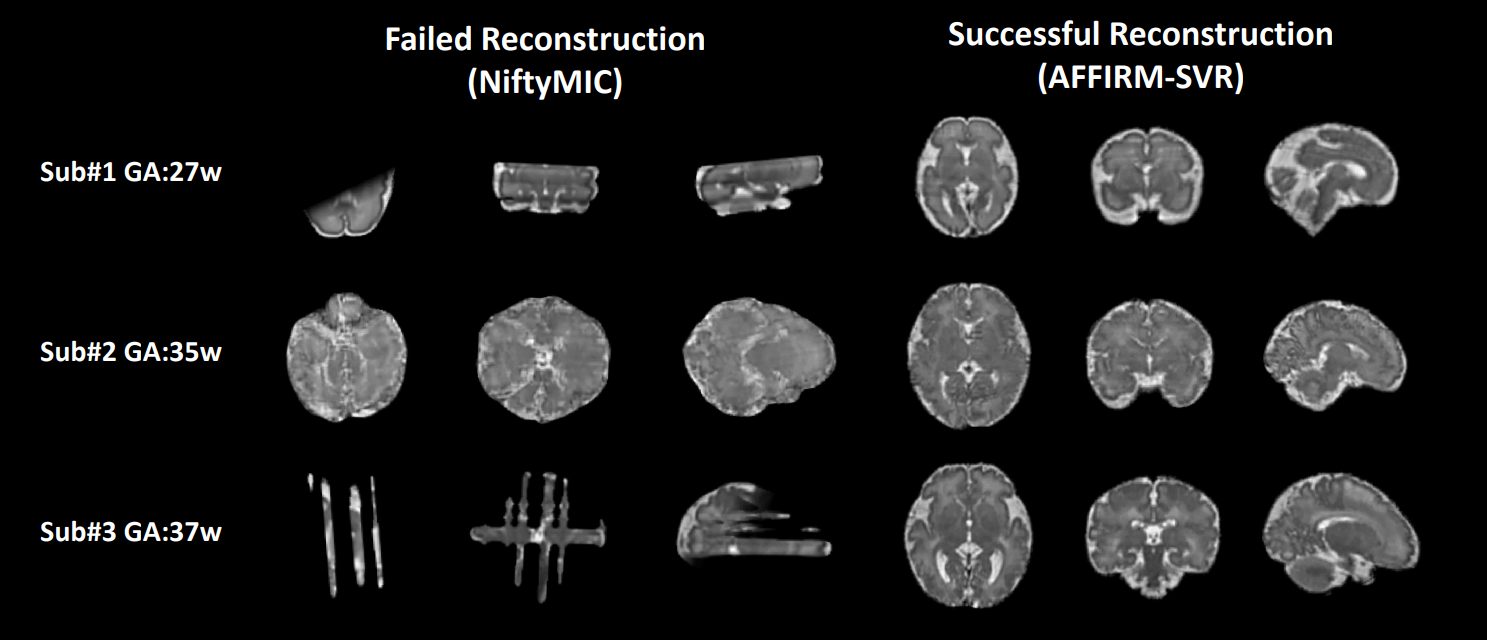}}
\caption{Super-resolution reconstruction of three fetal brains at different gestational ages (GA) that failed NiftyMIC pipeline (left) and were rescued using the AFFIRM-SVR method (right).}
\label{fig1}
\end{figure}

\section{Discussion}
Robust motion correction for multi-slice fetal brain MRI is a well-known challenge. In this work, we present a novel learning-based recursive framework, i.e., AFFIRM, that mimics the non-causal iterative optimization between motion estimation and volumetric reconstruction. Experiments on the simulated motion-corrupted data demonstrated the superior performance of the proposed AFFIRM compared to other learning-based methods and improvement of SRR of the fetal brain. Furthermore, as validated by real-world multi-slice MRI data, our method expanded the capture range of the SVR and increased the success rate of SRR. The fully automated coarse-to-fine framework can be easily implemented as a retrospective motion correction tool that may benefit fetal brain MRI research. 

Some of the learning-based motion estimation techniques, e.g., SVRnet and DeepPE, predict the fetal poses using only information from single slices. Though computationally efficient, they do not make good use of the sequential information which encodes the fetal poses in the consecutive slices. We, therefore, proposed a bi-directional recurrent network architecture to utilize the features from the present and adjacent slices for motion estimation. The results in Table.I indicate that the RNN-based networks, i.e., DeepPMT and AFFIRM, both improved the accuracy of fetal pose estimation. However, all the existing fetal brain estimation networks simply used one stack from a single orientation to predict the 3D poses in the canonical space. In other words, these networks essentially learn the fetal anatomical patterns using CNN-based feature extraction, and therefore, the predictions heavily depend on the training data distribution. It in turn will be relatively insensitive to small motion and may not handle the changing fetal brain anatomy over a large GA span. Many works also disregard the border slices because it is highly difficult to learn the patterns where the image features are sparse [15], [20], [31]. To overcome these problems, AFFIRM has a 3D branch that mimics the SVR scheme. The 3D volume in the canonical space is fed into the network as a reference to help improve the motion estimation by bridging it to the 2D slices. The CNNs here serve as data-driven feature detectors, which resemble the first step of SVR. However, it remains difficult to explicitly match the features with different dimensions in the deep network. Although this scenario can be considered an extreme case of 3D-3D feature matching, the imbalanced information impedes the fitting of the 3D network and increases the computation. Alternatively, we used feature fusion to learn the relationship between 2D and 3D features. Late fusion is a simple and predominant approach in multimodal learning [53]–[55], which, however, was not effective and even reduced the network performance according to Table II, possibly because the features deep in the 2D stream are too high-level to match the redundant 3D features using concatenation. Hence, we applied a specific fusion that explicitly learns the 2D and 3D spatial correspondence in the feature domain using an affinity function at the intermediate stage of the network, which significantly improved the network performance compared to late fusion. Furthermore, the recursive architecture also increased the estimation accuracy as the reference volumes were refined in each iteration.

Although AFFIRM achieved desirable performance of motion estimation regardless of fetal brain orientations and slice locations, it is still not sufficient to completely remove all motions by itself, especially small rotation and displacement. Furthermore, as shown in Fig. 5, we found that the motion at small GA remained hard to correct as the fetal brain is relatively small with only a few slices available. On the other hand, conventional SVR implemented in NiftyMIC performs well for small motion but easily fails for large motion due to limited capture range [56], [57]. The AFFIRM essentially increases the capture range by learning a prior fetal brain position in the canonical space and fine-tuning with respect to the given reference volume. Therefore, AFFIRM and SVR complement each other and can be integrated to handle both large and small motions (Fig. 9). 

Here, we trained the network on the simulated motion-corrupted data that was generated by a random procedure, similar to previous studies [15]. The simulated motion was controlled by a reasonable scale, i.e., an instantaneous angular velocity limit of up to 15 degrees per second in the 3D space, and the pose orientations also covered all possible poses in the 3D space. The network may further benefit from training on realistic motion patterns which are difficult to record directly and there are no established fetal head motion models to the best of our knowledge. Nevertheless, we speculate the network would predict well on the real data if generalized on a larger set of the simulated motion that includes the real motion trajectories of the fetal brain. The assumption was supported by the real-world experiments in Section III.E.  

The present work has several limitations. Above all, AFFIRM is essentially based on absolute pose estimation in the canonical space that is still scene-specific and requires adjustments to new scenarios, such as different image contrasts. Image harmonization and specific data augmentation may address the problem. Besides, AFFIRM can only handle inter-slice motion while a few slices will still be corrupted by in-plane motion. A general motion correction method that takes all types of motion into account will rescue more slices and further improve the entire pipeline.  Additionally, some acquisition parameters, e.g., slice thickness and slice ordering, are fixed in the network, which requires adjustment to configure on other datasets. Also, our network was trained on the normal fetal brains with only mild anomalies, e.g., mild ventriculomegaly and mega cisterna magna, and thus, the performance of the network on fetal brains with severe abnormalities is unknown. Future work will focus on encoding relative position and we will involve abnormal fetal brains and multi-center datasets for more robust generalization. 

\section{Conclusion}
In this work, we proposed an affinity fusion-based framework for random motion correction of multi-slice fetal brain MRI. The recurrent network learns the sequential information from multiple stacks of 2D slices and mimics the process of iterative SVR and volumetric reconstruction using affinity fusion with a scattered data approximation module. The proposed network outperformed the other state-of-the-art learning-based techniques for fetal brain motion estimation and improved the SRR results on the large motion-corrupted data compared to registration-only methods. This network can also be added to the conventional pipeline to initialize the multi-resolution SVR to address both large- and small-scale motions and improve the success rate of SRR. In summary, our learning-based motion correction scheme is an effective technique to handle arbitrary brain motion and help enhance the performance of the existing fetal brain MRI processing pipeline.

\end{document}